\documentclass[aps,prl,twocolumn,groupedaddress,showpacs,showkeys]{revtex4}%
\usepackage{graphics}% 
\begin{document}%
%----------------------------------------------------------------
% Definitions needed for the heading
%-------------------------------------------------------------------------
%\def\Barcelo{Barcel\'o}
\def\E{{\mathbf E}}
\def\n{{\mathbf n}}
%-------------------------------------------------------------------------
\title{Where is the electrostatic self-energy localized in general relativity?}
%-------------------------------------------------------------------------
\author{Carlos Barcel\'o}
\email[]{carlos@iaa.es}
%\homepage[]{http://www.iaa.es/}
%\thanks{Supported by the Spanish Ministry of Science and Technology}
%\altaffiliation{}
\affiliation{Instituto de Astrof\'{\i}sica de Andaluc\'{\i}a, CSIC,
Glorieta de la Astronom\'ia, 18008 Granada, Spain}
%-------------------------------------------------------------------------
%-------------------------------------------------------------------------
\author{Jos\'e Luis Jaramillo}
\email[]{Jose-Luis.Jaramillo@aei.mpg.de}
%\homepage[]{http://www.iaa.es/}
%\thanks{}
%\altaffiliation{}
\affiliation{
Max-Planck-Institut f{\"u}r Gravitationsphysik, Albert Einstein
Institut, Am M\"uhlenberg 1 D-14476 Potsdam Germany
}
%-------------------------------------------------------------------------
\date{2 December 2011; file new-charged-geometry.tex --- version 0.00; 
\LaTeX-ed \today}
%-------------------------------------------------------------------------
\bigskip
%-------------------------------------------------------------------------
\begin{abstract}
%-------------------------------------------------------------------------
\bigskip

We discuss an alternative way of prescribing the spacetime geometry associated with a non-radiating distribution of charged matter. It is based on the possibility that the electrostatic self-energy does not reside on the Coulombian field but in a matter pressure term of electromagnetic origin localized at the sources. We work out completely the well controlled spherically symmetric case, questioning the realization of Reissner-Nordstr\"om geometry in nature. Finally, we sketch an experiment that could distinguish between the standard and the alternative scenario.

%-------------------------------------------------------------------------
\end{abstract}
%-------------------------------------------------------------------------
%-------------------------------------------------------------------------
%-------------------------------------------------------------------------
\maketitle
%-------------------------------------------------------------------------

%----------------------------------------------------------------
% Local defines
%----------------------------------------------------------------

%\def\d{{\mathrm{d}}}
\def\e{{\mathrm e}}%
\def\g{{\mbox{\sl g}}}%
\def\Box{\nabla^2}%
\def\d{{\mathrm d}}%
\def\R{{\rm I\!R}}%
%--------------------------------------------------
\def\ie{{\em i.e.\/}}%
\def\eg{{\em e.g.\/}}%
\def\etc{{\em etc.\/}}%
\def\etal{{\em et al.\/}}%
%--------------------------------------------------
%----------------------------------------------------------------
\def\HRULE{{\bigskip\hrule\bigskip}}
%----------------------------------------------------------------

{\em Introduction.} A basic tenet of general relativity is that all forms of energy gravitate and do so in an equivalent manner. However, the concrete way in which this principle is realized depends on the precise spatial localization of the considered form of energy. In particular, this applies to the electrostatic self-energy. In its standard treatment the electromagnetic self-energies reside in the field. This naturally promotes in general relativity to the use of an electromagnetic stress-energy-momentum tensor built from the electromagnetic strength field as a source in Einstein equations. In particular, this leads to a description of the exterior of a spherical distribution of charged matter in terms of the Reissner-Nordstr\"om geometry~\cite{rn}. However, the way in which the electrostatic self-energy appears in these solutions contrasts with the uncharged gravitostatic case: whereas in the electromagnetic case the self-energy is localized in the field, in the pure gravitational case a natural interpretation is that it is localized within the gravitating object itself. This asymmetry leads us to discuss here the possibility according to which the electrostatic self-energy is actually localized in the charged object rather than in the field. Let us stress that this remark applies only to the electrostatic self-energy and not to the electromagnetic dynamical (radiative) degrees of freedom; in a sense, the effect of the electrostatic self-energy would be to renormalize the properties of matter at the source. Ultimately, the actual location of the self-energy can only be decided by experiment. It is one of the aims of this letter to highlight the plausibility of addressing this question experimentally.

\vspace{0.1cm}
{\em The mass formula for a dust shell.} In general relativity the external geometry surrounding a spherically-symmetric distribution of non-charged matter is described by the Schwarzschild solution:
\begin{equation}%
ds^2=-F(r)dt^2 +F(r)^{-1}dr^2 +r^2 d\Omega_2^2~,
\label{RN-solution}%
\end{equation}%
with $F(r)=(1-2M/r)$.
As a simple model for the matter distribution let us consider a matter shell made of dust. Using Lanczos-Israel matching conditions~\cite{lanczos,israel-thin-shells} (see also e.g. the treatment of thin shells in~\cite{visser-book}) one can paste an exterior Schwarzschild solution with an internal Minkowski spacetime through a thin-shell of variable radius $a(\tau)$ (this precise calculation can be found in~\cite{israel-thin-shells}). Here $\tau$ represents the proper time as measured by an observer attached to the shell. The matching conditions result in an equation for the
energy density in the shell
\begin{equation}
\rho=
{1\over 4\pi a}  
\left(\sqrt{1+\dot a^2} -\sqrt{F(a) + \dot a^2} \right)~,
\label{matching1}
\end{equation}
plus the conservation equation
\begin{equation}
{\d\over\d \tau} (\rho a^2)=-p {\d\over\d \tau} (a^2)~.
\label{conservation}
\end{equation}
Here $p$ represents the tangential pressure inside the shell, that we set to zero consistently with the
dust nature. In this case we can define a constant mass parameter $M_0$ as $M_0=4\pi a^2 \rho$. Then, isolating $M$ in~(\ref{matching1}) we obtain
\begin{eqnarray}
M=\sqrt{1 + \dot a^2} M_0-{M_0^2  \over 2a}~,
\end{eqnarray}
or using instead the derivatives with respect to the time $T$ associated with internal Minkowski observers,
\begin{eqnarray}
a'\equiv {da \over dT}~; ~~~~~\dot a =  a' {d T \over d \tau} = a' \left(1- a'^2 \right)^{-1/2}~,
\end{eqnarray}
we can express it as
\begin{eqnarray}
M= {M_0 \over \sqrt{1 - a'^2}} -{M_0^2 \over 2a}~.
\label{dust-shell}
\end{eqnarray}
This exact general relativity formula has a remarkably intuitive interpretation: The asymptotic mass (energy) of the configuration has a positive contribution coming from the special-relativistic mass of the moving matter (the mass of the constituents when infinitely separated and at rest, together with its kinetic energy which is encoded in the Lorentz factor) and a negative contribution coming from the gravitational self-energy of the rest mass.

{\em The mass formula for a charged dust shell.} The standard treatment of the electrostatic self-energy uses the electromagnetic field $F_{ab}$ and the energy-momentum tensor
$T_{ab}= {1 \over 4\pi} \left(F_{ac} F_b^{\;c} - 
{1 \over 4}g_{ab} F_{cd}F^{cd}\right)$.
Then, the geometry exterior to a spherical distribution of charged-matter is given by the Reissner-Nordstr\"om line element (\ref{RN-solution}), with $F(r)=\left(1-{2M \over r}+ {Q^2 \over r^2}\right)$
where $Q$ is the charge.

Repeating the thin-shell calculation above but for the charged-dust case, one obtains~\cite{kuchar}:
\begin{eqnarray}
M= {M_0 \over \sqrt{1 - a'^2}}  - {M_0^2 \over 2a} + {Q^2 \over 2a}~.
\label{energy-equation-charge}
\end{eqnarray}
The only difference with the non-charged case is the appearance of an electrostatic self-energy term. 

\vspace{0.1cm}
{\em Operational difference between mass and charge.} The central observation here is that whereas the electrostatic self-energy is localized in the field outside the matter distribution, the gravitostatic self-energy acts as if it were located in the shell itself. 

This point can be illustrated, for instance, by considering quasi-local masses in both situations. 
Although there does not exist a well-defined local notion of gravitational energy, consistent quasi-local notions can still be introduced when considering additional structures or certain symmetries~\cite{szabados}. In the spherically symmetric case here considered, there exists a canonical notion of quasi-local energy given by the Misner-Sharp mass~\cite{misner,szabados}. The evaluation 
of this quasi-local mass for a particular radius in the Schwarzschild geometry, leads to
%(see eq. (1.11) in~\cite{misner}) 
its constancy outside the shell: $M_{\rm QL}(r)=M$. A natural interpretation is that outside the shell there is no energy content contributing to the asymptotic mass of the configuration (whereas other quasi-local masses can lead to different interpretations in the pure gravitostatic case, see the recent discussion in~\cite{Frauendiener:2011rm}, we find that this one offers the clearest perspective on the gravitostatic-electrostatic asymmetry). 
Operationally this quasi-local mass can be retrieved by measuring geometric quantities, for example, the convergence effect on light rays at different radii.

The situation concerning the quasi-local mass is completely different when considering the charged Reissner-Norsdtr\"om geometry. The quasi-local mass of the exterior geometry is no longer constant. This mass is now
\begin{eqnarray}
\label{e:Komar_conservation}
M_{\rm QL}(r)= M - {Q^2 \over 2r}~,
\end{eqnarray}
an increases with $r$. The behavior of the quasi-local mass fulfills the idea that the electrostatic self-energy is localized on the electric field surrounding the charge (the charged shell, in the studied case) extending up to infinity. Therefore, the external geometry is not empty but possesses an energy field which contributes to the quasi-local mass more and more as one encompasses more and more of the field. This fact also entails that one could distinguish a charged from a non-charged distribution of matter by performing exclusively gravitational measurements: their associated geometries are different. 

From this perspective, there is an asymmetry in the treatment of electrostatic and gravitostatic self-energies in standard Einstein-Maxwell theory. Given the similar description of the static gravitational and electric field distributions  
from a Newtonian perspective (namely, through a Coulombian field), this discrepancy in the standard relativistic treatment is a remarkable feature. In the next section we will discuss an alternative way of dealing with charged distributions of matter such that this asymmetry disappears. Before doing so, let us review the reasons behind the standard assumption that the electrostatic self-energy resides in the field.

First, it is a standard exercise in electrostatics to prove that the work one has to do against the electric interaction to put a set of point charges together is equivalent 
to the volume integral of an electric field energy density.
Second, the existence of solutions of Maxwell's equations representing electromagnetic waves
also offers evidence in favor of a energy-field point of view. A radiating system loses energy and reciprocally a receptor heats up by absorbing radiation. This is a strong indication that electromagnetic waves carry energy. It is an almost inescapable fact that one has to associate some energy density to the field of an electromagnetic wave, given the immense body of experiments we have (let us just mention that within a theory of action-at-a-distance the previous assertion might be questioned~\cite{feymann-wheeler}). In this letter we assume that this is the case.

 However, without resorting to experiments involving gravity, it is not at all obvious that one has to associate an energy-density to a non-radiative field configuration. The formal relation 
between the work done to set up a static configuration and the volume integral of its electrostatic-field energy density could be precisely that, {\em just a formal relation}.
More specifically,  the electrostatic self-energy 
in a region ${\cal R}$ can be expressed in two equivalent ways
\begin{eqnarray}
\label{e:energydensities}
E = {1 \over 8\pi} \int_{\cal R} \!\!\! |D\phi|^2 dV= 
{1 \over 8\pi} \int_{\cal R} \!\!\! \rho_c \phi \; dV
- {1 \over 8\pi} \int_{\partial {\cal R}} \!\!\!\! \phi E_\perp \; dS \ .
\end{eqnarray}
The first form of $E$ suggests that the energy density is localized at the
field, whereas the second expression involves (up to a boundary term,
$E_i=-D_i\phi$) an energy density
localized at matter. When integrating in a volume ${\cal R}$, such a 
difference in the interpretations of the density is only formal
since they are related by the Poisson equation, $\Delta \phi = -\rho_{c}$,
upon integration by parts.
However, the situation changes when switching on gravity in a relativistic setting:
Einstein equation is a local equation in which the sources are precisely energy densities. 
Therefore in the passage to general relativity 
the particular chosen form for the electrostatic energy density %in (\ref{e:energydensities})
directly impacts the gravitational field dynamics.
It is remarkable that, in the standard Einstein-Maxwell treatment,
(\ref{e:energydensities}) finds a general relativistic analogue in stationary situations, 
but precisely only as long as we consider regions without matter.
More concretely, the mass quantity
\begin{eqnarray}
\label{e:GRstationary_mass}
\!\!\!
M({\cal S})= 
%{-1 \over 8\pi} \!\! \left[
{1 \over 4\pi} \!\!
\oint_{\cal S} \!\!\nabla_a t_b dS^{ab}
 \! + {1 \over 4\pi} \!\! \oint_{\cal S}\!\! \left( \! A_c t^c E_{\perp} \!+\! B_c t^c B_{\perp} \! \right) dS
%\right]
\end{eqnarray}
takes the same value on two 2-surfaces ${\cal S}$ and ${\cal S}'$ as long
as their enclosed volume ${\cal R}$ does not contain  charged matter
(cf. \cite{Simon:1984qb} for a related more general discussion).
The first term is the standard Komar mass, absent in the non-gravitational 
expression (\ref{e:energydensities}),
whereas the second term corresponds (for vanishing magnetic field $B_{\perp}=0$)
to the boundary term in (\ref{e:energydensities}), 
taking into account surface orientation and field pressure 
[$t^a$ is the timelike 
Killing vector associated with stationarity, and $A_a, B_a$ are 4-vector potentials 
for $F_{ab}$ and its dual ${}^*F_{ab}$, the latter  defined only in vacuum]. 
The breakdown of the  analogy when trying
to incorporate the matter term in (\ref{e:energydensities}) into the general relativistic 
expression (\ref{e:GRstationary_mass}), indeed reflects the specific choice for
the electrostatic energy density in the standard treatment of Einstein-Maxwell theory, namely
as associated with the field distribution.
 However, to the best of our knowledge, there is not experimental evidence of whether or not the Coulomb field gravitates as if it really represented a distribution of field energy.

Such an experimental gap, together with the argued asymmetry in the treatments of electrostatic and gravitostatic self-energies in Einstein-Maxwell theory and the formal properties of conserved quantities in stationarity,
lead us to propose an alternative way of taking into account the electrostatic self-energy in general relativity. 

\vspace{0.1cm}
{\em Alternative energy-momentum tensor: A pressure of electromagnetic origin.} Motivated by the pure gravitational case in which the gravitostatic self-energy is localized in the shell, the alternative proposal to treat the electrostatic self-energy is the following: 
To incorporate the electrostatic self-energy by modifying the matter energy-momentum tensor through the introduction of a pressure term of electromagnetic origin.

Applying a thermodynamical argument to the work to be done against the electrostatic repulsion to vary the shell radius, $dE= -pdA$, and then writing 
\begin{eqnarray}
d \left({Q^2 \over 2 r}\right) =- {Q^2 \over 2r^2} dr = -p d (4\pi r^2) ~,
\end{eqnarray}
we identify the following electromagnetic pressure form
\begin{eqnarray}
p= {Q^2 \over 16\pi~r^3}~. 
\end{eqnarray}
This pressure takes into account the electric repulsion of the different (equal sign) charges within the shell.

Now, given a shell with an energy density and a pressure of this form one can repeat the analysis involving Eqs. (\ref{matching1}) and (\ref{conservation}). This leads to a modification of Eq. (\ref{dust-shell})
\begin{eqnarray}
\label{e:mass_new}
\!\!\!\!\!M= { 1 \over \sqrt{1 - a'^2}} M_Q-{M_Q^2 \over 2a}~, \ \ \hbox{with} \ \ M_Q  =  M_0 + {Q^2 \over 2a}~.
\end{eqnarray}
This formula has a natural interpretation. The total asymptotic energy of the configuration, $M$, has a special relativistic contribution $\gamma M_{Q}$, with $\gamma$ the Lorentz factor, plus a negative gravitostatic self-energy contribution $-M_{Q}^2/2a$. The mass $M_{Q}$ is the sum of the rest mass $M_0$ of the constituents of the shell when completely separated plus its electrostatic self-energy $Q^2/2a$. 

When comparing this mass formula with the standard relativistic formula (\ref{energy-equation-charge}) based in Einstein-Maxwell theory, we notice that the new formula has additional terms taking into account the inclusion of the positive electrostatic self-energy both into the gravitational self-interaction part and into the kinematical mass affected by the Lorentz factor. In a sense, this offers a better matching with the spirit of the equivalence principle of general relativity, putting on equal footing all forms of energy. Note that it is not only the mass formula what changes in this scenario, but more importantly the very geometry of spacetime. In this description, the external geometry of the charged distribution of matter is of Schwarzschild type. In particular, this implies that no pure gravitational experiment could tell between a charged and a non-charged spherical distribution of matter. As we mention, this is in contrast with the standard Reissner-Nordstr\"om scenario.  

\vspace{0.1cm}
{\em Some properties of the new proposal.} Let us consider the situation in which the charged dust shell is lowered to a certain position $a_0$ by using external means (e.g. using rigid strings attached to the shell through which one can extract energy from the system or imprint energy on it). Its total energy when released, that is kept constant by the free dynamics, would be [cf. Eq. (\ref{e:mass_new}) with $a'=0$]
\begin{eqnarray}
M= \left(M_0+{Q^2 \over 2a_0}\right)
\left[1- {1 \over 2a_0}\left(M_0+{Q^2 \over 2a_0}\right)\right]~.
\label{static-energy}
\end{eqnarray}
This leads to a positive $M$ for all $a_0>a_e$ where
\begin{eqnarray}
a_e \equiv {1 \over 4}
\left(
M_0 + \sqrt{M_0^2+4Q^2}
\right)~.
\end{eqnarray}
Notice that the Schwarzschild radius $r_M \equiv 2M$ is always smaller than $a_0$ for all $a_0\in (a_e,\infty)$ except for the special critical radius $a_0=a_c$
\begin{eqnarray}
a_c\equiv {1 \over 2} \left(M_0+\sqrt{M_0^2+2Q^2}\right)~,
\end{eqnarray}
for which $r_M=a_c$. Thus, except for this critical configuration, all the other  configurations are consistent with the shell matching procedure applied in this paper, which assumes a timelike trajectory for the shell.

From expression (\ref{static-energy}) it follows that there exist equilibrium configurations where the electric repulsion exactly compensates the gravitational attraction. This equilibrium configurations satisfy $dM/da_0=0$. For $M_0>0$ this equation has only one physically possible solution
\begin{eqnarray}
a_s= {Q^2 \over 2(Q^2-M_0^2)} \left(2M_0 + \sqrt{M_0^2+3Q^2}\right)~.
\end{eqnarray}
In order $a_s$ to acquire a well defined positive value, it is necessary that $M_0<|Q|$. Then, the equilibrium position $a_s$ is always larger than the Schwarzschild radius.
However, by calculating $d^2M/da_0^2$ one can check that all these equilibrium configurations are unstable: if perturbed, either they expand towards infinity or they contract to form a Schwarzschild black hole. Given a particular situation with $M_0<|Q|$, the position $a_s$ separates an external region in which the electric repulsion wins from an internal region where the gravitational attraction dominates. In the case in which $M_0>|Q|$ there are not equilibrium configurations, all of them have the tendency to collapse towards black-hole formation.

We make two observations at this point.  First, these configurations are able to implement Wheeler's idea of matter without matter. One could construct solutions with $M_0=0$ but $M \neq 0$ (in the equilibrium configuration $M=2\sqrt{3}|Q|/9$). Then, the asymptotic mass is the resultant of the accumulation only of electrostatic energy.  

Secondly, let us define a charge length scale as $r_Q \equiv 2|Q|$. In terms of the Reissner-Norsdtr\"om geometry, the condition $r_M=r_Q$ marks the so-called extremal limit. For $r_M<r_Q$ the black hole limit of the Reissner-Norsdtr\"om geometries contain naked singularities. In the present alternative scenario however, the exterior geometry is always of Schwarzschild type so that, independently of the values of $r_M$ and $r_Q$, the black hole limit of these solutions never develops naked singularities. For instance, the previous equilibrium configurations all have $r_Q>r_M$. Moreover, by using external means one could build configurations with an asymptotic-mass to charge ratio well beyond the extremal limit without any special complication. This would be the case if, with the help of some rigid strings, one lowers the charged dust shell up to a radial position $a_0 \gtrsim a_e$. For $M_0<|Q|$ it is not difficult to check that $a_e<a_s$ so that positions $a_0$ close enough to $a_e$ always corresponds to places in which the gravitational attraction wins once the shell is released. Therefore, once the position $a_0 \gtrsim a_e$ is reached, one can leave the shell free to collapse to form a black hole. This black hole will have an arbitrarily small mass $M$ for the given charge. Now, recall that elementary particles are all well beyond extremality. For instance, an electron has a ratio $r_Q/r_M \sim 10^{21}$. Therefore, as opposed to the standard treatment, this framework is capable of accommodating simple classical models for elementary particles.

\vspace{0.1cm}
{\em Discussion of experimental feasibility.} In this letter we are highlighting that there is no experimental evidence of the actual realization of Reissner-Nordstr\"om geometry in nature. In fact, the previous discussion suggests that it could be the case that the geometry associated with a spherically symmetric distribution of charged matter were, after all, Schwarzschild geometry. This would imply that by performing only gravitational experiments one would not be able to distinguish whether a lump of matter is charged or not. 

However, what could be clearly distinguished by performing a gravitational experiment is, indeed, whether the electrostatic self-energy associated with the Coulomb interaction is localized in the field or within the body itself. The simplest experiment one can think of would be to look at whether the Newtonian force between charged bodies has or has not corrections with respect to the square law. The idea would be to design an experiment to look for Newtonian forces at small distances (see e.g.~\cite{gravity-experiments}) but now with a charged drive mass.   
    
Let us put some conservative figures based on state of the art experiments. In the Newtonian limit
and at the particular length scale of the experiment $r_S$, 
one would need the (repulsive) gravitational force $mc^2 r_Q^2/4r^3$ of purely electrostatic origin (derived from the Reissner-Nordstr\"om potential) to be comparable to the standard (attractive) Newtonian force $-mc^2 r_M/r^2$. Otherwise the former would be completely masked by the latter.
In such a case, $r_Q \sim \sqrt{r_M r_S}$.  Now, typical drive masses in these experiments are of the order $10^{-6}{\rm Kgr}$ ($r_M \sim 10^{-21} {\rm m}$) and $10^{-3}{\rm m}$ is already a well-controlled length scale. In these circumstances one would need $r_Q \sim 10^{-12}{\rm m}$ or, what is the same, a charge of the order of
\begin{eqnarray}
Q \sim 10^{5}~{\rm Coulombs}~.
\end{eqnarray}
This is quite a large charge (equivalent to $10^{24}$ electrons) for being accumulated in such a small volume, but it does not seem utterly impossible for experiments in a reasonable near future. 
  
{\em Acknowledgements.} We would like to thank A. Harte and W. Simon for
illuminating certain formal discussions.

%------------------------------------------------------------------------------
 
%------------------------------------------------------------------------------
\end{document}